\documentclass[groupedaddress,twocolumn,prl]{revtex4}
\usepackage{epsfig,amsfonts,amsmath}
\newcommand{\be}{\begin{equation}}
\newcommand{\ee}{\end{equation}}
\newcommand{\ba}{\begin{eqnarray}}
\newcommand{\ea}{\end{eqnarray}}
\def\ket#1{\left\vert #1 \right\rangle}
\def\bra#1{\left\langle #1 \right\vert}

\def\bea{\begin{eqnarray}}
\def\eea{\end{eqnarray}}
\def\ben{\begin{eqnarray*}}
\def\een{\end{eqnarray*}}

\def\>{\rangle}

\def\<{\langle}

\def\l{\left}

\def\r{\right}

\newcommand{\fig}[1]{Fig.~\ref{fig:#1}}

\newcommand{\Hp}{H_{\mathrm{pair}}}
\newcommand{\Up}{U_{\mathrm{pair}}}
\newcommand{\Vp}{V_{\mathrm{pair}}}
\begin{document}

\title{Limitations of Quantum Simulation Examined by Simulating a
Pairing Hamiltonian using Nuclear Magnetic Resonance}

\author{Kenneth R. Brown, Robert J. Clark, and Isaac L. Chuang}

\affiliation{Center for Bits and Atoms, Research Laboratory of
Electronics, \& Department of Physics\\ Massachusetts Institute of
Technology, Cambridge, Massachusetts 02139, USA}

\date{\today}

\begin{abstract}

Quantum simulation uses a well-known quantum system to predict the
behavior of another quantum system.  Certain limitations in this
technique arise, however, when applied to specific problems, as we
demonstrate with a theoretical and experimental study of an algorithm
to find the low-lying spectrum of a Hamiltonian.  While the number of
elementary quantum gates required does scale polynomially with the
size of the system, it increases inversely to the desired error bound
$\epsilon$.  Making such simulations robust to decoherence using
fault-tolerance constructs requires an additional factor of $\sim 1/\epsilon$
gates. These constraints are illustrated by using a three qubit
nuclear magnetic resonance system to simulate a pairing Hamiltonian,
following the algorithm proposed by Wu, Byrd, and Lidar \cite{Wu:02}.

\end{abstract}


\maketitle

The unknown properties and dynamics of a given quantum system can
often be studied by using a well-known and controllable quantum system to mimic the behavior of the original
system.  This technique of {\em quantum simulation} is one of the
fundamental motivations for the study of quantum computation
\cite{Fey:82,Lloyd:96,AbLloyd:99}, and is particularly of interest
because a quantum simulation may be performed using space and time
resources comparable to the original system.  Such ``efficient''
scaling is dramatically better than the exponentially large resource
requirements to simulate any general quantum system with a classical
computer, as Feynman originally observed \cite{Fey:82}.

Recent work has continued to arouse great interest in quantum
simulation, because it offers the possibility of solving
computationally hard problems without requiring the resources
necessary for algorithms such as factoring \cite{Shor:97} and
searching \cite{Grover:97a}.  Experimental results have demonstrated
simulations of a truncated oscillator and of a three-body interaction
Hamiltonian, using a nuclear magnetic resonance (NMR) quantum
computer \cite{Cory:99,Cory:00}, and explored various solid-state
models on two qubit systems \cite{Suter:04,Laflamme:04,Cory:04,Du:04}.
Interest has also extended to simulating complex condensed matter
systems with quantum optical systems \cite{Jane:02}, demonstrated
vividly by the observation of a superfluid to Mott insulator
transition in a Bose-Einstein condensate \cite{Greiner:02}.

Often overlooked in the discussion of quantum simulations, however, is
the question of desired precision (or error $\epsilon$) in the final
measurement results. Current quantum simulation
techniques generally scale poorly with desired precision; they demand
an amount of space or time which increases as $1/\epsilon$, broadly
translating into a number of quantum gates which grows exponentially
with the desired number of bits in the final answer. Why is this
scaling behavior so poor, and what is its physical origin?

Consider as a specific example the problem of calculating the energy
gap $\Delta$ between the ground state $\ket{G}$ and the first excited
state $\ket{E_1}$, of a Hamiltonian $H$. $\Delta$ can be found using
the following steps: 1) map the Hilbert space of the system to be
simulated to $n$ qubits, 2) prepare the computer in the state
$\ket{\Psi_I}=c_G\ket{G}+c_E\ket{E_1}$, 3) evolve under the
Hamiltonian for times $t_i$, 4) extract the phase difference as a
function of time between the evolution of the ground and first excited
state.

Two methods for calculating the phase difference, and thus $\Delta$,
are as follows. The first method uses the phase estimation algorithm
\cite{Cleve:98,Nielsen:book}. This method relies on the quantum
Fourier transform (QFT) and requires simulating the Hamiltonian for
times $t_k= 2^k t_0$, for integer $k$ from $0$ to $q$. Since the input
state is a superposition of $\ket{G}$ and $\ket{E_1}$, the measured
phase will either be $E_G t_0$ or $(E_G+\Delta)t_0$, where $E_G$ is the
ground state energy. One needs to run the algorithm on average
$1/c_E^2$ times to get both values and thus measure $\Delta$.

The second method does not use the QFT, and instead simulates the
Hamiltonian for times $t_k=k t_0$, for integer $k$ from 0 to $Q$, and
then measures any operator $M$ such that $\bra{G}M\ket{E_1}\neq
0$. Typically, any operator that does not commute with the Hamiltonian
suffices.  After calculating $\langle M(t_k)\rangle$, one classically
Fourier transforms (FT) over the averaged values yielding a spectrum
$\langle M(\omega) \rangle$ with peaks at $\pm\Delta$ and $0$.

For fixed precision, obtaining $\Delta$ up to error $\epsilon$ for
fixed $\epsilon$, both methods can be ``efficient,'' in that the
number of elementary steps (or quantum gates) required increases only
polynomially with the number of qubits $n$, if the Hamiltonian can be
efficiently simulated and the initial states efficiently prepared.  A
$d$-qubit Hamiltonian can be simulated with a number of gates of order
$O(n^d)$ assuming two qubit interactions between any
qubits\cite{AbLloyd:99}. If one assumes only nearest neighbor two
qubit gates, it scales as $O(n^{d+1})$. Most physical systems of
interest are described by two-body interactions which can be described
by four qubit Hamiltonians. 

Consequently, the challenge of designing efficient quantum simulations
is choosing a property that can be efficiently extracted.  However, no
general measurement method is known which allows $\Delta$ to be
measured efficiently {\em with respect to the precision} using such
quantum simulations.  For error $\epsilon$, the number of digits of
precision in the result is $\log(1/\epsilon)$, and both of the above
methods require $\sim 1/\epsilon$ steps (or gates) to obtain this
precision.  In contrast, an efficient algorithm would only require a
number of steps polynomial in $\log(1/\epsilon)$. The origin of this
limitation lies not only in the inability to design efficient
measurements, but also in the accumulation of errors which occurs in
the course of performing a quantum simulation.

Here, we consider these limitations on the precision of results
obtained by quantum simulations in the context of a specific algorithm
for the simulation of pairing models, as proposed by Wu, Byrd and
Lidar (WBL)\cite{Wu:02}, which follows the framework of the two
methods described above.  We present a study of the errors in its
discrete time step implementation, and experimental results from a
realization using a 3 qubit nuclear magnetic resonance (NMR) quantum
computer, answering three questions: 1) What are the theoretical
bounds on the precision of the quantum simulation?  2) How do faulty
controls affect the accuracy of a simulation? 3) Can the theoretical
bounds on precision be saturated by an NMR implementation?

The WBL algorithm uses the classical FT algorithm described above to
solve the question of the low-lying energy gap in pairing
Hamiltonians. Pairing Hamiltonians are used to describe both nuclear
dynamics and superconductivity \cite{Song:book,Mahan:book,Ring:book}
and are usually written in terms of Fermionic creation and
annihilation operators $c^\dagger$ and $c$ as
\be
H_{\mathrm{pair}} = \sum_{m=1}^n\frac{\epsilon_m}{2} (c^\dagger_m c_m
+ c^\dagger_{-m} c_{-m}) + \sum_{m,l=1}^N
V_{ml}(c^\dagger_mc^\dagger_{-m}c_lc_{-l}) 
\,,
\nonumber
\ee
where $n$ is the total number of modes, $\epsilon_m$ is the onsite
energy of a pair in mode $m$, and $V_{ml}$ are the coupling constants
between modes.

WBL map the pairing Hamiltonian onto the qubit Hamiltonian
\begin{equation}
	\Hp = \sum_{m=1}^n \frac{\nu_m}{2}\l( -Z_m\r) +
	\sum_{m<l} \frac{V_{ml}}{2}(X_mX_l + Y_mY_l) 
\,,
\nonumber
\end{equation}
where $X_m, Y_m,$ and $Z_m$ are the Pauli operators on the $m$th qubit
and $\nu_m=\epsilon_m+V_{ll}$ (dropping an unimportant global energy
shift, and using the standard convention $Z\ket{0}=\ket{0}$). The
number of modes that can be simulated equals the number of qubits $n$,
and the number of pairs equals the total number of qubits in the state
$\ket{1}$.  WBL show that for a specific number of pairs, one can
approximately prepare the state $\ket{\Psi_I}$ by quasiadiabatic
evolution.  Since $\Hp$ is a $2$-body Hamiltonian, the system's
evolution can be efficiently simulated on a quantum computer for any
number of qubits \cite{Lloyd:96}.  WBL propose implementation of their
algorithm using an NMR quantum computer, in which the operator $M$ is
simply $Z$ for a single spin. An advantage of the ensemble nature of
NMR is that a single measurement for a simulated time $t$ yields
$\langle M(t)\rangle$. Fixing a maximum energy width and desired
precision makes the FT independent of the number of qubits.

Let us begin by addressing the first question posed above, regarding
theoretical bounds on the precision of this quantum simulation: how
does the number of gates scale with the error $\epsilon$? The WBL
method requires constructing an operator that approximates the
simulated Hamiltonian for times $t_k$.  The classical FT then yields
an error of $2\pi E_{max}/Q$ where $E_{max}$ is the largest detectable
energy $1/t_0$ $(\hbar = 1)$. In the case of using phase-estimation
and the QFT, setting $2^q$=$Q$ yields the same precision. How long
does it take to implement the Hamiltonian for a time $Q t_0$ compared
to implementing a Hamiltonian for time $t_0$? In general, the operator
is assumed to be constructed of repetitions of the basic time step and
requires $Q$ more gates or time. This leads to the number of gates
scaling inversely with the error.  A similar problem faced in quantum
factoring is overcome in Shor's algorithm by a clever way to perform
the modular exponentiation \cite{Shor:97}.

A second bound on the number of gates required arises in calculating
the time required to perform the algorithm.  Quantum simulations
typically employ a Trotter formula to approximate a Hamiltonian from
combinations of non-commuting Hamiltonians\cite{Nielsen:book}.  For
example, given the ability to evolve under Hamiltonians $H_A$ and
$H_B$, one can approximate evolution under $H_A+H_B$ with bounded
error.  To lowest order, $\exp\l(-it(H_A+H_B)\r) = \l(\exp\l(-itH_A/k
\r) \exp\l(-itH_B/k \r) \r)^k+\delta$, where for $\|[H_A,H_B]\| t^2\ll
1$, the error $\delta$ is $O(t^2/k)$. Higher-order techniques can
yield an error $O(t^{m+1}/k^m)$ at the cost of needing $O(2^m)$ more
gates \cite{Suzuki:92}.

This approximation method leads to a subtle but important difficulty
in reducing the gate count for simulations.  It is apparent that the
Trotter formula demands an exponential increase in the number of
discrete gates for an exponential decrease in the error.  However, from
a Hamiltonian control perspective, this conclusion seems unfair,
because the total {\em time} required can be small even if the gate
count is high.  Specifically, the gate $U_A(t/k)=\exp\l(-itH_A/k\r)$
requires $1/k$ the time needed to implement $U_A(t)$. Therefore, the
simple Trotter method given above requires only time $2t$, independent
of $k$. This implies that ``Trotterization'' errors involved in
approximating desired Hamiltonians can be reduced efficiently with respect
to the time cost.

Unfortunately, this optimistic observation is incompatible with fault tolerant error correction \cite{Preskill:97a,Gottesman:97a}, which will likely be needed to extend simulation times beyond limits imposed by qubit decoherence times. This is because the fault-tolerant implementation of $U_A(t/k)$ takes approximately the same amount of time as the gate $U_A(t)$, whether using teleportation \cite{Gottesman:99a} or the Solvay-Kitaev approximation \cite{KSV}.

Consequently, fault-tolerant simulations using the Trotter formula and
the FT/QFT require a number of gates and amount of time that scales as
$1/\epsilon^2$.  Circumventing this problem would require removing the
inefficiency of Trotterization, and constructing methods to
approximate $U_H(t)$ with error $\epsilon$ using
${\rm poly}(\log(1/\epsilon))$ gates. However, such methods would imply that
the approximation of $U_H(qt)$ could take only ${\rm poly}(\log(q))$ more
gates than the simulation of $U_H(t)$. Such a dramatic simplification
may hold for specific $H(t)$, but is unlikely to be possible for
general $H(t)$.  The WBL algorithm studied here unfortunately does not
scale efficiently when made fault tolerant.

These theoretical bounds establish that present quantum simulations
such as the WBL algorithm, using the QFT or the FT, require a number
of gates which scales inversely with the desired answer precision for
two reasons: fault-tolerant gate construction and the precision of a
finite FT.  Therefore, the time required for a $d$-qubit quantum
simulation is $O(n^{d}/\epsilon^r)$, where $r\geq 1$ varies depending
on the approximation methods employed, and $r=1$ when quantum error
correction and fault-tolerant gates are not used.

We turn now to the second question, which concerns the impact of faulty controls in
a real physical implementation of the WBL algorithm.  Recall that the
foundation of the WBL algorithm is approximation of the unitary
evolution under $\Hp$, $\Up(qt_0)=\exp\l( -i\Hp qt_0\r)$.  An ideal
NMR implementation accomplishes this by a repeatable pulse sequence
$\Vp(t_0)$, where $\Up(qt_0)\approx \left(\Vp(t_0)\right)^q$. $\Hp$
contains three noncommuting parts: $H_0=\sum_m\frac{\nu_m}{2} (-Z_m)$,
$H_{XX}=\sum_{m<l} \frac{V_{ml}}{2} X_mX_l$, and $H_{YY}=\sum_{m<l}
\frac{V_{ml}}{2} Y_mY_l$. Assuming that the corresponding unitary
operators $U_0(t), U_{XX}(t),$ and $U_{YY}(t)$ can be implemented,
$\Vp(t_0)$ can be constructed using the third order Trotter-Suzuki
formula \cite{Trotter:58,Suzuki:92}
\begin{eqnarray}
	\Vp(t_0) &=& \left[ U_0(t_0/2k)U_{XX}(t_0/2k)U_{YY}(t_0/k) \times
		     \right.
	\nonumber 
\\ &&	\left. U_{XX}(t_0/2k)U_0(t_0/2k) \right]^k
	\,, 
\nonumber
\end{eqnarray} 
yielding an expected error $\|\Up(t_0)-\Vp(t_0)\|=O(t_0^3/k^2)$.

However, this ideal procedure is not actually achieved in a real
experiment because the unitaries are not direct implementations of the
Hamiltonians but instead composed from a series of pulses. These
pulses depend on assumptions about the system Hamiltonian that become unreasonable for short simulated times. The reason is that all real systems have small, often unknown,
energy shifts that are averaged away for large simulated times. When not using
a fault-tolerant construction these shifts can lead to faulty
controls. In atomic physics, for example, undesired Stark shifts need
to be carefully accounted for in order to get exact rotations
\cite{Wineland:98}.

Control errors in NMR quantum computation arise, for example, since
single qubit gates require finite time and unwanted two qubit coupling
occurs during this time.  In a static magnetic field $ B_0\hat{z}$,
the unitary evolution of a typical used spin
system\cite{Cory:04,Vandersypen:04} in the rotating frame is given by
$U_{ZZ}(t)=\exp\l( -i \sum_{ij} \frac{\pi}{2}J_{ij}Z_iZ_j t\r)$, where
the $J_{ij}$ are the scalar coupling constants.  The time $t_{\pi}$
required for a radiofrequency (RF) pulse to rotate individual spins by
$\pi$ radians is much smaller than the typical delay times $t_{d}$
during which no RF is applied, $t_{\pi} \ll t_{d} \approx 1/J_{ij}$.
Thus, it ordinarily suffices to approximate the RF pulses as
$\delta$-functions in time, implementing perfect single qubit
rotations $R^i_\phi(\theta)=\exp\l[
\frac{i\theta}{2}(X_i\cos\phi+Y_i\sin\phi)\r]$.  However, this
approximation breaks down as $t_d$ becomes comparable to $t_\pi$,
causing the expected evolution to be best described not by discrete
one and two-qubit gates, but instead by the piecewise continuous
time-dependent Hamiltonian $H_{nmr}(t)=\sum_i g_i(t)X_i +
\sum_if_i(t)Y_i + \sum_{i<j} \frac{\pi}{2}J_{ij}Z_iZ_j$.  This
discrepancy leads to additional errors in implementations of quantum
algorithms and simulations, which, for a small number of qubits, can
be mitigated using optimal control techniques\cite{Laflamme:04,
Khaneja:02}.

\begin{figure}
\begin{center}
\epsfig{file=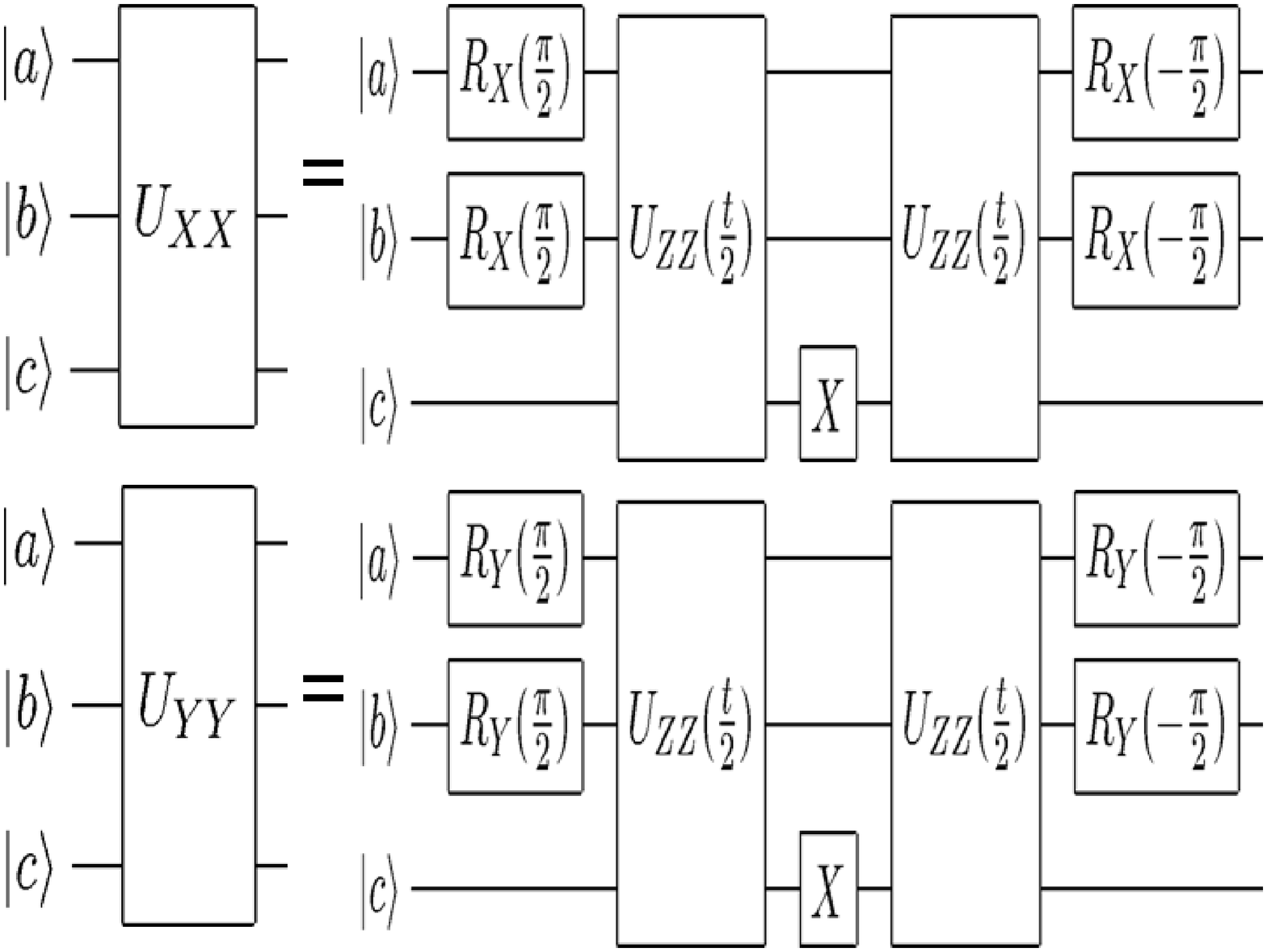, width=7cm, height=3.3cm}
\caption{Three qubit quantum circuits for the unitaries $U_{XX}$ (top)
and $U_{YY}$ (bottom), implemented using method $W1$.  These are
depicted for simulating Hamiltonian $H_2$ (see text).  For $H_1$,
$\frac{\pi}{2}$ pulses are applied to qubit $c$ in parallel with those
on $a$ and $b$, and the decoupling $X$ pulse is
omitted. \label{fig:circuit}}
\end{center}
\end{figure}

The impact of such control errors in an NMR implementation of the WBL
algorithm can be studied by comparing a baseline realization with no
control error compensation (denoted $W1$) versus another with simple
error compensation (denoted $W2$).  The baseline $W1$ realization
implements $U_0$ using composite pulses to create rotations about the
$\hat{z}$ axis, $U_0=\prod_m R^m_{\pi/2}(\pi/2) R^m_0(\pi\nu_m)
R^m_{-\pi/2}(\pi/2)$; an equivalent method, used
elsewhere\cite{Wu:02,Laflamme:04} temporarily shifts the rotating
frame.  Control errors arise in the simulation of $U_{XX}$ and
$U_{YY}$, which are generated by applying single qubit pulses to
rotate the scalar coupling from the $\hat{z}$ axis to the $\hat{x}$
and $\hat{y}$ axis, using the quantum circuit in \fig{circuit}.
Control errors in this baseline realization are thus small only when
delays needed to generate $U_{XX}$ are long compared to the time
required to perform single qubit gates, but also short enough that the
Trotter error is small.

A simple, scalable compensation technique for control errors provides
a contrasting realization of the WBL algorithm for comparison. This
$W2$ realization accounts for unwanted two qubit coupling during
single qubit gates by reducing delay times during which coupling is
desired.  Specifically, every instance of $R_{\phi_1}(\theta_1)
U_{ZZ}(t) R_{\phi_2}(\theta_2)$ is replaced with $R_{\phi_1}(\theta_1)
U_{ZZ}(t-\alpha) R_{\phi_2}(\theta_2)$, where
$\alpha=\frac{t_\pi}{2\pi}(\theta_1+\theta_2)$.  This technique was
critical in the successful implementation of Shor's algorithm with
NMR\cite{Vandersypen:01}; here, it is used with care, since many
Hamiltonians have the same $\Delta$ as the pairing Hamiltonian of
interest, and it is possible to tune $\alpha$ to get the right
$\Delta$ for the wrong reasons.

\begin{figure}
\begin{center}
\begin{tabular}{c}
\epsfig{file=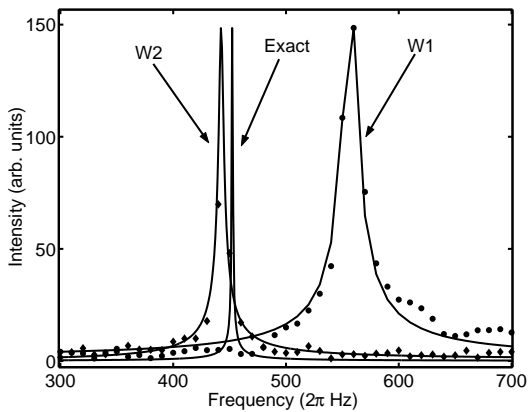, width=7cm}
\end{tabular}
\caption{Frequency-domain spectra of Hamiltonian $H_2$ obtained using
methods $W1$ (marked by circles) and $W2$ (diamonds). Dots are
experimental data, and solid lines are Fourier transforms of an
exponentially decaying sinusoid (with four free parameters) fit to the
time dependent data. The width of the exact curve is taken to be the
dephasing rate ($1/\tau$) of the $^{13}\mathrm{C}$
nucleus. \label{fig:h2}}
\end{center}
\end{figure}

Numerical simulations comparing $W1$ and $W2$ show that the effect of
such control errors on the WBL algorithm is a shift in the estimated
gap value $\Delta$ from the expected value.  This shift can be quite
significant, as shown in \fig{h2}, and indeed can dominate errors due
to other imperfections, such as the Fourier transform.  Compensating
for unwanted scalar couplings in NMR implementations of quantum
simulations is thus vital for obtaining correct results;
implementations with other physical systems will similarly have to
deal with faulty controls.  

Finally, we consider the third question: saturation of the predicted
precision bounds with an NMR implementation of the WBL algorithm.  The
WBL algorithm is parameterized by the number of qubits, $n$, the
simulated time step $t_0$, the number of steps $Q$, the degree of
Trotterization $k$, the adiabatic time step $t_{ad}$, and the number
of adiabatic steps $S$.  An NMR system is characterized by a
characteristic decoherence time $\tau$, and following the discussion
above, it is convenient to work with a small number of qubits for
times shorter than $\sim 3\tau$, such that quantum error correction is
unnecessary. We implement two specific instances of the pairing
Hamiltonian $\Hp$ involving three modes ($n=3$) and two pairs, leading
to a $3$-dimensional Hilbert space spanned by $\ket{101}$,
$\ket{110}$, and $\ket{011}$. The simulation is started the ground
state of the two spin up subspace of $H_0$, $\ket{011}$, prepared
using temporal labeling \cite{Steffen:99}.

WBL estimated the expected size of the system that could be simulated
without error correction by choosing $k/t_0 = 0.1 \Delta$ and
$\epsilon_{FT}=\Delta$. They found that the number of gates required
scales as $3 n^4\Delta/\epsilon_{FT}$, including the necessary
decoupling pulses. The gate time, $t_g$ is assumed equal to $10^{-5}
\tau$ and for up to $n=10$, $\Delta$ can be found to precision
$\epsilon \approx\Delta$.  Here, we find $\Delta$ to precision
$\epsilon \approx\Delta/100$, and the number of qubits is $n=3$,
consistent with the WBL bound $n\leq 4$ for these parameter choices.

The first stage of the WBL algorithm is to quasiadiabatically evolve
into the ground state $\ket{\Psi_I}$ of $\Hp$, with discrete changes
in the simulated Hamiltonian, using a procedure previously
demonstrated \cite{maxcut}.  The Hamiltonian used at each discrete
step $s$ is $H_{ad}(s)=(1-s/S)H_0+(s/S)\Hp$, where $S$ is the maximum
number of steps. Unitary evolution $U_{ad}$ at each step for time
$t_{ad}$ is then approximated using the above pulse sequences, as
$U_{ad}=\prod_{s=0}^S V_{ad}(s,t_{ad})$.  Preparation of the state
$\ket{\Psi_I}$ requires evolving at a rate faster than that for
adiabatic evolution, thereby exciting the state $\ket{E_1}$. This
quasiadiabatic evolution is accomplished by reducing $S$ or $t_{ad}$
compared to the adiabatic case \cite{Wu:02}. Higher-energy states will
also be excited, but $S$ and $t_{ad}$ can be adjusted to minimize
this. Quasiadiabatic evolution in this experiment was attained with
$S=4$ steps and $t_{ad}=1/700$ s.  Note that for $\|H_{XX}+H_{YY}\|
\gg \|H_0\|$ there can be a phase transition as $s$ is changed
\cite{McKenzie:96}; as the gap goes to zero at the phase transition
this can be problematic, since the number of steps required for
successful quasiadiabatic evolution grows inversely with the gap.

The second stage of the algorithm is evolution of the state
$\ket{\Psi_I}$ under the pairing Hamiltonian for $Q$ timesteps of
duration $t_0$ with $k=2$.  These parameters are chosen such that
$1/(Q t_0) \approx \Delta/100$ , $Qt_0<\tau$, and $k/t_0 > 0.1\Delta$.
Note that many $Q$ and $t_0$ yield the same $\epsilon_{FT}$; this is
used to our advantage below.

We performed our experiments using a $500$ MHz Varian
$^{\mathrm{UNITY}}${\it INOVA} spectrometer and $^{13}$C-labeled
CHFBr$_2$, with coupling strengths $J_{HC}=224$ Hz, $J_{HF}=50$ Hz,
and $J_{CF}=-311$ Hz.  The two pairing Hamiltonians simulated were
$H_1$, the ``natural'' Hamiltonian, in which $V_{12}=\pi J_{HC}$,
$V_{13}=\pi J_{HF}$, and $V_{23}=\pi J_{CF}$, and a harder case, an
artificially constructed Hamiltonian $H_2$, in which $V_{ab}=\pi
J_{HC}$ and $V_{ac}=V_{bc}=0$. For both Hamiltonians, $\nu_1=150\pi$
Hz, $\nu_2=100\pi$ Hz, and $\nu_3= 50\pi$ Hz. Implementation of $H_2$
required an additional decoupling pulse compared to $H_1$, following
the circuit in \fig{circuit}.

Each Hamiltonian was simulated for times $t_0$ to $Qt_0$, using the
$W1$ and $W2$ methods, and an NMR spectrum was acquired for each time
duration.  A classical discrete FT of the NMR peak intensities of one
spin (hydrogen) over the $t_k$ yielded four spectra of $\Hp$. The
experimental result $\Delta_{exp}$ was determined by a least-squares
fit of the highest-signal NMR peak to a damped sinusoidal function
with frequency $\Delta$ and decay rate $\tau_e$.

Ideally, the result should find $\Delta=218 \cdot 2\pi$ Hz for $H_1$,
and $\Delta=452 \cdot 2\pi$ Hz for $H_2$, as determined by direct
diagonalization.  Note that for $H_2$, $\Delta$ is the energy
difference between $\ket{G}$ and $\ket{E_2}$, since $\ket{E_1}$ is not
connected by usual adiabatic evolution; the larger gap requires that
$t_0/k$ be smaller when simulating $H_2$.  For the experimental
result, we expect that $\Delta_{exp}$ =$\Delta +\epsilon_{sys} \pm
\epsilon_{FT}$ where $\epsilon_{sys}$ is an offset due to
Trotterization and/or faulty controls.  $Q$ and $t_0$ determine the
theoretical bound on the precision, in the absence of control errors,
and the experiment should saturate this bound when the $\Delta_{exp}$
is $\epsilon_{FT}$ from the actual value $\Delta$.

\begin{table} 
\begin{center}
\begin{tabular}{|c|c|c||c|c|c|c|}
\hline
Model & $\Delta$/Hz & Method & $\Delta_{exp}$/$2\pi\cdot$Hz &
$\tau_e$/ms & $t_0$/ms & $Q$\\ 
\hline
$H_1$& 218 $\cdot 2\pi$ & $W1$ & $227\pm 2$ & 180 & 1 & 400\\
     &                  & $W1$ & $220\pm 2$ & 250 & 2 & 200\\
\hline
$H_2$& 452 $\cdot 2\pi$ & $W1$ & $554\pm 10$ & 30 & .5 & 200\\
     &     & $W2$ & $440\pm 5$  & 80 & .5 & 200\\
%     & 452 & $F$  & $400\pm 50$ & 4 & .2 & 90\\ method ``F'' vetted from paper
\hline
\end{tabular}
\end{center}
\caption{Experimental results for gaps found for Hamiltonians $H_1$
and $H_2$. Estimated gaps ($\Delta_{exp}$) and effective coherence
times ($\tau_{e}$) for given time steps $t_0$ and number of steps $Q$
are obtained by least-squares fitting of the time-dependent NMR peaks
to an exponentially-decaying sinusoid.}
\label{table}
\small 
\normalsize
\end{table}

Experimental results for the spectra of $H_2$ are shown in \fig{h2};
experimental parameters $Q$ and $t_0$ and numerical results
from the analysis for $\Delta_{exp}$ and  $\tau_e$ for each experiment are
summarized in Table I.

The impact of systematic and random errors was investigated by
simulating $H_1$ with $W1$ (no control error compensation) for
$\epsilon_{FT}=2.5 \cdot 2\pi$ Hz at two different simulation times, $t_0= 1$ ms and
$t_0=2$ ms.  As expected, the random error for both cases is $\approx
\epsilon_{FT}$. Note that the systematic error increases with smaller
$t_0$. This signals that the error due to unwanted scalar coupling
becomes larger than the errors due to the Trotter
approximation. Consequently, a slightly longer $t_0$ yields a systematic
error that is within $\epsilon_{FT}$ of the exact answer, saturating
the predicted theoretical bounds on precision.

Convergence to the correct result is another important issue for all
discrete time simulations.  For this $3$-qubit system, we performed a
detailed numerical simulation to determine that $t_0=2$ was
optimal. For a large system this is no longer possible, and
convergence tests would need to be used to verify the answer. The
procedure would reduce $t_0$ (or increase $k$) until the change in
$\Delta_{exp}$ was smaller than the desired precision.

While the results for Hamiltonian $H_1$ were good even without control
error compensation, the effects of control errors were very evident in
the results for $H_2$.  Hamiltonian $H_2$ was implemented with $W1$
(no error compensation) and $W2$ (simple error compensation) for
$\epsilon_{FT}=10 \cdot 2\pi$ Hz and $t_0=0.5$ ms. The shorter time step was
necessary because the larger $\Delta$ made the simulation more
sensitive to Trotter errors. Comparing the $W2$ and $W1$ results shows
that with no control error compensation, a gap $\Delta$ is found that
is $\Delta/5$ away from the actual value.  In contrast, with simple
error compensation $\Delta$ is $\epsilon_{FT}$ from the actual value,
saturating the theoretical bound.  Future implementations should
certainly strive to detect and bound control errors; this could be
done by verifying that $\Delta_{exp}$ scales as $t_0^3$ for small
values of $t_0$, as theoretically expected.

In conclusion, we have studied the theoretical and empirical bounds on
the precision of results obtained with quantum simulations, in the
context of the pairing Hamiltonian algorithm proposed by Wu, Byrd, and
Lidar. We have implemented the smallest problem
instance that requires quasiadiabatic evolution, verifying that the algorithm computes the gap $\Delta$ to within the
precision of the method.  We also find, however, that simulations of
this type are particularly sensitive to systematic errors in the
applied Hamiltonian and that fault-tolerant implementations are {\em
inefficient} with respect to precision using current Trotter
approximation methods.

Nevertheless, in practice, when only limited precision is desired and
for a sufficiently large system, quantum simulations may still
outperform classical numerical simulation, as demonstrated for
molecular energies \cite{Head-Gordon:05}.  Avoiding the cost of
precision is desirable, and can be done by designing quantum
simulations to explore questions that are insensitive to the
microscopic details of the Hamiltonian \cite{Milburn:05}. How to
develop quantum simulations for faulty small scale (10-20 qubit)
quantum computers that can outperform classical computations remains
an open question.

\bibliographystyle{apsrev}
% \bibliography{refs}

\end{document}